\documentclass{aa}
\usepackage{epsfig}
\begin{document}

\thesaurus{08(02.01.2; 09.10.1; 08.05.2; 13.25.5; 13.09.6; 08.09.2)}

\title{Infrared photometry and spectroscopy of the supersoft X-ray 
source RX J0019.8+2156 (= QR And)}

\author{H. Quaintrell\inst{1}\fnmsep\inst{2} \and
R. P. Fender\inst{1}\fnmsep\inst{3}}

\institute{Astronomy Centre, C.P.E.S., University of Sussex, Falmer,
  Brighton, BN1 9QJ, U.K.  \and Department of Physics, The Open
  University, Walton Hall, Milton Keynes MK7 6AA, U.K. \and
  Astronomical Institute ``Anton Pannekoek'' and Center for
  High-Energy Astrophysics, University of Amsterdam, \\
  Kruislaan 403, 1098 SJ Amsterdam, The Netherlands.}

\offprints{H.Quaintrell@open.ac.uk} 

\date{Received 27 January 1998 / Accepted 7 April 1998}

\authorrunning{H. Quaintrell \& R. P. Fender}

\titlerunning{Infrared observations of RX J0019.8+2156}

\maketitle 

\begin{abstract}
  
  We present JHK photometry and spectroscopy of RX J0019.8+2156.  The
  spectrum appears to be dominated by the accretion disc to at least
  2.4 $\mu$m, over any other source of emission.  We find Paschen,
  Brackett and He II lines strongly in emission, but no He I.  There
  are satellite lines approximately 850km s$^{-1}$ either side of the
  strongest, unblended hydrogen lines. These satellite lines may be
  the spectral signature of jets from the accretion disc.


\end{abstract}

\section{Introduction}
\label{sec:intro}

The `supersoft' X-ray sources form a new and exciting but
inhomogeneous class of X-ray objects. Though a few members of the
class were known before ROSAT, it is this observatory which has
characterised the class as objects with temperatures of 15 -- 80 eV
and luminosities approaching the Eddington limit for a solar mass
object, i.e.  $10^{36}$ -- $10^{38}$ erg s$^{-1}$. The model which
best fits the observations has a white dwarf steadily or cyclically
burning hydrogen-rich material accreted onto its surface at $\sim
10^{-7}$ M$_{\sun}$ yr$^{-1}$ (van den Heuvel et al.
\cite{vdheuvel92}).

van den Heuvel et al. (\cite{vdheuvel92}) theorise that this material
may be supplied by a near main-sequence A/F companion star, more
massive than the white dwarf and in a close binary orbit, which is
overflowing its Roche lobe.  Such systems are known as close binary
supersoft sources (CBSS), for further discussion of the nature of the
class see Kahabka \& van den Heuvel (\cite{kahabka97}).

35 supersoft X-ray sources have so far been identified within our
galaxy, M31, the Local Group galaxy NGC 55, the LMC \& SMC (listed in
Kahabka \& van den Heuvel \cite{kahabka97}). Of these systems
\object{RX J0019.8+2156} (hereafter RX J0019) is the only northern
hemisphere CBSS identified, so far, in our galaxy.  RX J0019 is a
close binary with a 15$\fh$85 period and both short-term and long-term
optical variability, with orbital variability of $\Delta V \sim 0.5$
(Beuermann et al. \cite{beuermann95}, hereafter B95; Greiner \& Wenzel
\cite{greiner95}). UBVRI photometry at photometric maximum and minimum
reveal a very blue continuum, in common with other CBSS.  The optical
and UV spectra show Balmer $\alpha$ and He II lines, especially
$\lambda\lambda$1640,4686\AA, in emission with enhanced blue wings,
but no He I lines (B95).  Balmer $\beta$ and $\gamma$ lines exhibited
P-Cygni type profiles.

\section{Observations}
\label{sec:obs}

\begin{table}
    \caption{Spectroscopic and photometric JHK magnitudes and spectral 
      indices. JHK fluxes were obtained from spectra by fitting the 
      continua with low order polynomials and letting the flux be the 
      value of the fit at 1.25, 1.6 and 2.2 $\mu$m. The photometric
      phases were calculated using the ephemeris of Will \& Barwig 
      (\cite{will96}).
      The spectral index, $\alpha$, was calculated for spectra 
      dereddened by E(B--V)=0.1.}
    \begin{flushleft}
      \begin{tabular}{ccccccccc}
        \hline
        Band&\multicolumn{3}{c}{Photometry} & 
        &\multicolumn{3}{c}{Spectroscopy} \\
        \cline{2-4}\cline{6-8}
        &$\phi_{phot}$&mag&mJy&&$\phi_{phot}$&mJy&$\alpha$\\
        \hline
        J&0.35&12.00&24.09&&0.89&14.7&0.82\\
        H&0.35&11.97&15.97&&0.91& 9.6&--\\
        K&0.36&11.83&11.49&&0.97& 7.6&0.85\\
        \hline
      \end{tabular}
    \end{flushleft}
    \label{tab:RXJflux}
\end{table}

We have used the United Kingdom Infrared Telescope (UKIRT) to make
photometric and spectroscopic infrared observations of RX J0019, in
service mode.

\subsection{Photometry}
\label{sec:obs:phot}

\begin{figure}
  \epsfig{file=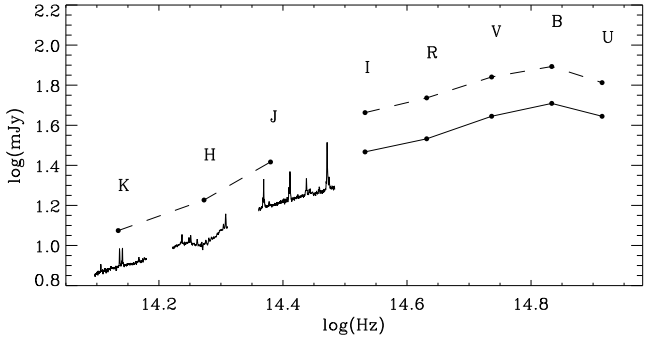}
  \caption{Dereddened optical--infrared  energy distribution of 
    RX J0019. Solid and dashed lines correspond to photometric
    minimum and maximum, respectively. UBVRI data are from B95.}
  \label{fig:phot}
\end{figure}

The photometric observations were carried out 1994 December 14, with
the IRCAM3 array operating in 0$\,^{\prime\prime}\!\!\!$.$\,$286 per
pixel mode.  Integration times ranged from 5 -- 20 s.  SA 114--750 was
used as the photometric standard assuming K=12.02, J--K=--0.09 and
H--K=--0.04.  The results are given in table \ref{tab:RXJflux} and
plotted, after dereddening by E(B-V)=0.1 (B95), in fig.
\ref{fig:phot}. Errors are typically 5\%.  These photometric
observations were originally presented in Fender \& Bell~Burnell
(1996).

\subsection{Spectroscopy}
\label{sec:obs:spec}

JHK spectra of RX J0019 were taken by CGS4 using the 75 lines/mm
grating, on 1997 August 13.  To produce each stellar spectrum a series
of short exposures were taken, alternated with equal length exposures
of a piece of nearby blank sky.  Krypton, Xenon and Argon lamp spectra
were taken in order to perform wavelength calibration.  Observations
of a nearby bright star, HR133 (K=6.65, J--K=0.02, H--K=0.01), were
taken in order to remove atmospheric spectral features and to flux
calibrate the observations.

UKIRT staff flat fielded the target and standard star observations,
subtracted the sky exposures from the corresponding stellar
observations, and then averaged these sky subtracted images to produce
a {\it reduced group} of observations. From the reduced groups we
optimally extracted (Horne \cite{horne86}) the stellar spectra and
differences in exposure times were corrected for.

The arc spectra were then extracted, the lines identified and
dispersion relations found by fitting low order polynomials. The
dispersion calibrations were subsequently applied to the target and
standard star spectra. The J-band spectrum was taken using the second
order of the spectrograph and had 13.2\AA\ resolution. The H and
K-band spectra were taken with the first order of the spectrograph and
had 26.4\AA\ resolution.

The removal of atmospheric features from the target star spectra and
flux calibration were achieved by dividing by an A star spectrum,
whose hydrogen absorption lines had been interpolated across, and
multiplying by a blackbody (BB) spectrum. The BB spectrum was
calculated using the effective temperature of the A star
(T$_{\mathrm{eff}}$ was assumed to be 8810K for an A2IV star) and its
JHK magnitudes. However, it proved impossible to interpolate across
the hydrogen lines between $\sim$ 1.8 and 2.1 $\mu$m, where there are
a series of strong atmospheric bands.  Hydrogen lines in this region
of the target star were hence discarded as unreliable. Generally, the
fluxes of the CGS4 spectra should be correct to better than 20\%.

\begin{figure}
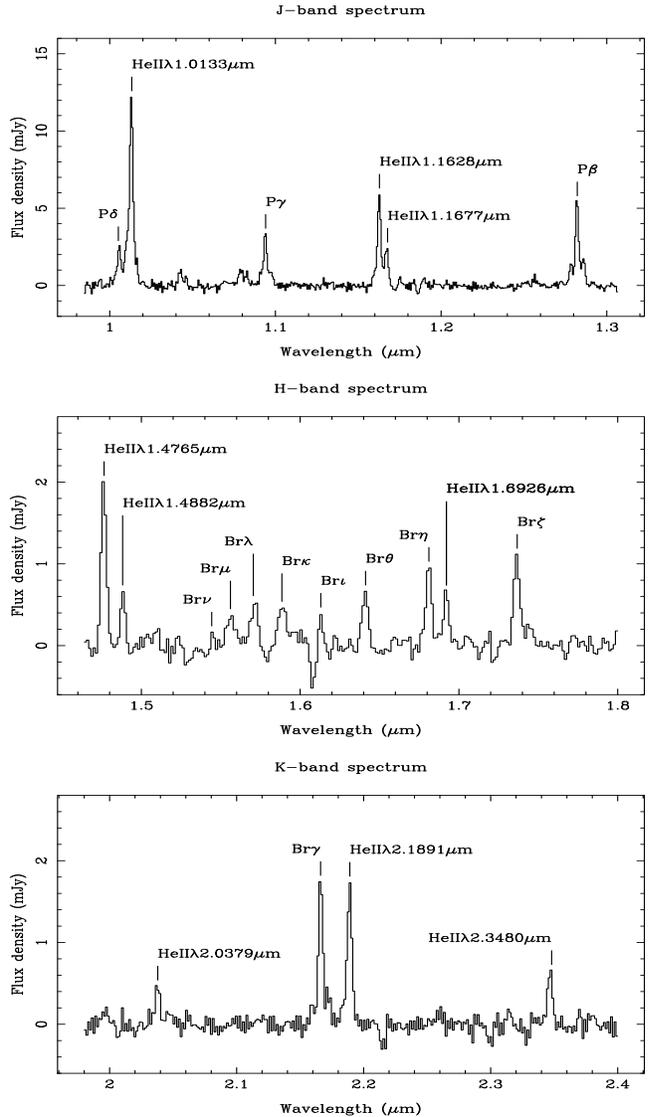

  \leavevmode
  \centering
  \epsfig{file=RXJ0019-J.eps,height=9.5cm,width=5cm,angle=270}
  \epsfig{file=RXJ0019-H.eps,height=9.5cm,width=5cm,angle=270}
  \epsfig{file=RXJ0019-K.eps,height=9.5cm,width=5cm,angle=270}
  \caption{JHK continuum subtracted RX J0019 CGS4 spectra, 
    showing hydrogen Paschen and Brackett, and He II lines in
    emission. No He I emission is observed.}
  \label{fig:0019spec}
\end{figure}

In addition to the infrared observations, the possible detection of a
jet (see below) motivated us to look for a radio counterpart. However,
approximately 2 hr of observations with the Ryle Telescope at 15 GHz
failed to detect a counterpart to a 3 $\sigma$ limit of 1.5 mJy
(G.G.~Pooley, private communication).

\section{Results}
\label{sec:rxj0019}

The infrared source was positively identified with the optical
counterpart indicated in Greiner \& Wenzel (\cite{greiner95}).  After
dereddening by E(B--V) = 0.1 mag (B95) and using the extinction
coefficients of Cardelli et al. (\cite{cardelli89}) our photometric
and spectroscopic data are plotted in fig. \ref{fig:phot}.  The
continua of the dereddened J and K spectra are reasonably flat, hence
we can measure the spectral indices for them, we find they are 0.82
and 0.85 respectively (see table \ref{tab:RXJflux}). The photometry
was taken around maximum light, whereas the spectroscopy was taken
around minimum light. From fig. \ref{fig:phot} it appears that the
difference in flux between maximum and minimum light in the infrared
is comparable to that seen in the optical.  Both the photometry and
spectroscopy hint at a dip in flux around the H band and/or an excess
of flux around J or K.

\begin{table}
  \caption{Fluxes and equivalent widths of unblended He II emission 
    lines, not contaminated by atmospheric features. Compare with 
    EW(He II$\lambda$4686\AA) $\sim$ 10\AA\ (B95).  The FWHM were 
    calculated by fitting single Gaussians.}
  \begin{flushleft}
    \begin{tabular}{ccccc}
      \hline
      $\lambda_{rest}$ &Flux& Flux density & EW    & FWHM\\ 
      ($\mu$m)&(10$^{-15}$erg s$^{-1}$)& (mJy)& (\AA) &(km s$^{-1}$)\\
      \hline
      1.4765 &11.1$\pm$2.2&0.72$\pm$0.15& 7.0$\pm$1.4&775$\pm$156\\
      1.4882 &2.9$\pm$1.9&0.19$\pm$0.13& 1.9$\pm$1.2&648$\pm$361\\
      1.6926 &3.4$\pm$0.9&0.25$\pm$0.06& 3.2$\pm$0.8&715$\pm$194\\
      2.0379 &2.0$\pm$0.8&0.18$\pm$0.07& 3.5$\pm$1.4&659$\pm$304\\
      2.1891 &5.5$\pm$0.7&0.52$\pm$0.07&11.4$\pm$1.5&637$\pm$80 \\
      2.3480 &1.8$\pm$0.7&0.20$\pm$0.07& 4.5$\pm$1.7&571$\pm$194\\
      \hline
    \end{tabular}
  \end{flushleft}
  \label{tab:heII}
\end{table}

\begin{figure}
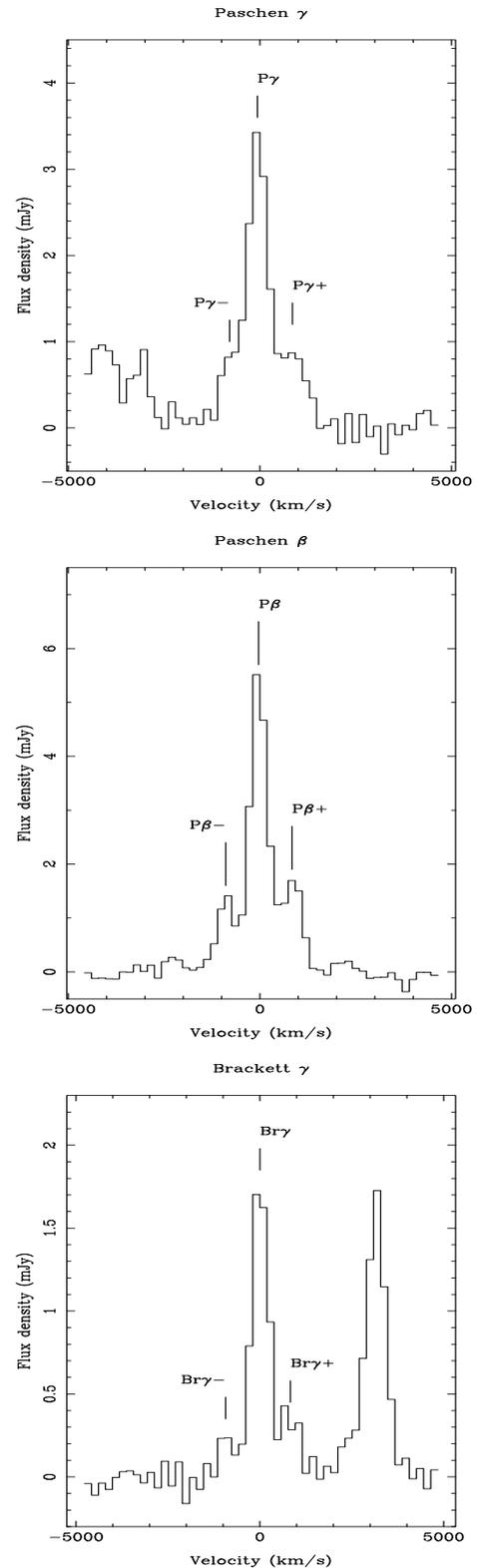

  \leavevmode
  \centering
  \epsfig{file=Pgamma.eps,width=7cm,height=7cm}
  \epsfig{file=Pbeta.eps,width=7cm,height=7cm}
  \epsfig{file=Bgamma.eps,width=7cm,height=7cm}
  \caption{HI lines in emission, continuum subtracted. Note the
           red- and blue-shifted components, labelled with + and
           -- signs respectively, which are signatures of high
           velocity components, possibly jets.}
  \label{fig:jets}
\end{figure}

The CGS4 spectra reveal the Paschen, Brackett and He II lines to be
strongly in emission (fig. \ref{fig:0019spec}). No He I emission or
absorption is observed. The equivalent width of He II
$\lambda$2.1891$\mu$m, found to be 11.4\AA, is comparable with the
value of $\sim$ 10\AA\ found for He II $\lambda$4686\AA\ by B95.  The
features of the infrared spectrum are therefore consistent with those
of the optical and UV.  The fluxes, flux densities, equivalent widths
and FWHM were measured for each unblended He II line and are recorded
in table \ref{tab:heII}. No photospheric absorption features, such as
CO bands, are observed in the spectra.

Close inspection of the three strongest, unblended HI lines observed
in the infrared spectra (Brackett $\gamma$, Paschen $\beta$ and
$\gamma$) reveal a strong emission line close to the rest wavelength
of the lines, and weaker emission features to either side (see fig.
\ref{fig:jets}). No other HI or He II lines correspond to the
wavelengths of the weaker emission lines. Although some of the
wavelengths correspond roughly with He I lines, there is no evidence
for He I emission elsewhere in either the infrared or optical spectrum
of RX J0019. In general, the red feature is noticeably stronger than
the blue. This structure does not seem to be present, at least to the
same degree, in the He II lines.

\section{Discussion}

\begin{table}
    \caption{Parameters of fits to HI lines using 3 Gaussians 
      per line, where velocities are relative to the laboratory 
      wavelengths.}
  \begin{flushleft}
    \begin{tabular}{cccc}
      \hline
      Spectral&$V_{peak}$&FWHM&Peak flux\\
      feature&(km s$^{-1}$)&(km s$^{-1}$)&(mJy)\\
      \hline
      P$\gamma$-- & --790 $\pm$ 197 & 542  $\pm$ 441 & 0.8 $\pm$ 0.4\\
      P$\gamma$  & --64  $\pm$ 52  & 577  $\pm$ 159 & 3.4 $\pm$ 0.5\\
      P$\gamma$+ & +850 $\pm$ 260 & 1026 $\pm$ 614 & 1.0 $\pm$ 0.3\\
      \hline
      P$\beta$--  & --891 $\pm$ 45  & 488  $\pm$ 114 & 1.4 $\pm$ 0.3\\
      P$\beta$   & --33  $\pm$ 14  & 525  $\pm$ 37  & 5.5 $\pm$ 0.3\\ 
      P$\beta$+  & +838 $\pm$ 44  & 606  $\pm$ 114 & 1.7 $\pm$ 0.2\\
      \hline
      Br$\gamma$--& --931 $\pm$ 163 & 386  $\pm$ 430 & 0.3 $\pm$ 0.2\\
      Br$\gamma$ & +2   $\pm$ 33  & 531  $\pm$ 85  & 1.8 $\pm$ 0.2\\
      Br$\gamma$+& +832 $\pm$ 167 & 556  $\pm$ 438 & 0.4 $\pm$ 0.2\\
      \hline
    \end{tabular}
  \end{flushleft}
  \label{tab:jets}
\end{table}

No photospheric absorption lines are seen in the infrared spectrum
taken at minimum light, indicating that either the underlying spectral
type of the donor is not later than F, we are seeing the system at low
inclination and irradiation is supressing absorption, or the accretion
disc is dominating the emission.  Through the optical and IR to at
least 2.4 $\mu$m the orbital modulation of flux appears to be nearly
constant, implying that they are dominated by the same component. The
continuum is flatter than that expected for a stellar photosphere,
suggesting the accretion disc is dominant.

Given the variability of the source over a variety of timescales
(Greiner \& Wenzel \cite{greiner95}), simultaneous multiwavelength
observations of RX J0019 would be necessary to properly determine the
relative contribution of light sources, and consequently place
constraints on the spectral class of the donor, inclination of and
distance to the system.

One explanation of the satellite lines to P$\gamma$, P$\beta$ and
Br$\gamma$ is that they are the infrared signature of jets, similar to
the Doppler shifted components of the Balmer $\alpha$, $\beta$ and He
II $\lambda$4686\AA\ lines of another supersoft source, \object{RX
  J0513.9-6951} (hereafter RX J0513, Crampton et al.
\cite{crampton96}; Southwell et al.  \cite{southwell96}), in which
$v_{\rm jet} \cos i \sim 3800$ km s$^{-1}$.  With this interpretation
in mind we fitted each of these three composite lines profiles with
three Gaussians. The details of these Gaussians are given in table
\ref{tab:jets}. The mean separation of the satellite lines from the
core is $844 \pm 76$ km s$^{-1}$, and their mean FWHM is $600 \pm 220$
km s$^{-1}$.  Assuming the jet interpretation is correct, and the jets
are perpendicular to an accretion disc which lies in the plane of the
binary, the true velocity of the outflow is simply a function of the
inclination angle ($v_{\rm jet} = (844 / \cos{i})$ km s$^{-1}$.).
Following Shahbaz et al. (\cite{shahbaz97}) we can also estimate the
opening angle of the jet, $\phi = \sin^{-1} (\Delta v/ v_{\rm
  obs}\tan{i})$, where $\Delta v$ is the FWHM of the satellite lines
and $v_{\rm obs}$ the observed velocity.

Meyer-Hofmeister et al. (\cite{meyer97}) model orbital modulation of
the system with a best-fit inclination of 56$^{\circ}$, corresponding
to a outflow velocity of $\sim 1500$ km s$^{-1}$ and an opening angle
of 28$^{\circ}$. Hachisu \& Kato (\cite{hachisu98}) model outbursts of
the source with an inclination of 45$^{\circ}$, corresponding to a
velocity of $\sim 1200$ km s$^{-1}$ and a very large opening angle of
45$^{\circ}$.  An inclination angle of around 75$^{\circ}$ would be
required to derive the same velocity as the jet in RX J0513, and would
imply a much more collimated jet ($\phi \sim 10^{\circ}$).
Alternatively the lower jet velocity could be intrinsic and reflect
the lower escape velocity required from a lower mass white dwarf.
Indeed Hachisu \& Kato (\cite{hachisu98}) argue for a mass of around
0.6 M$_{\odot}$ for the white dwarf in RX J0019, around half that
estimated for RX J0513 (Southwell et al. \cite{southwell96}).

{\em Note added in manuscript --} Just before submission we became
aware of a related work by Becker et al.  (\cite{becker98}). They have
found transient satellite lines around hydrogen and He II lines in
optical spectra of RX J0019. They also interpret the satellite lines
to be the signature of jets and find $v_{jet}\cos{i} \sim$ 800 km
s$^{-1}$, consistent with this work.

\begin{acknowledgements}
  We thank the staff responsible for the \\ UKIRT service program for
  their assistance in obtaining these observations. The UKIRT is
  operated by the Observatories on behalf of PPARC. The data were
  reduced using FIGARO software and computers provided by STARLINK.
  Thanks to T. R. Marsh for the use of his MOLLY software for spectral
  analysis. HQ acknowledges receipt of a PPARC studentship.  RPF was
  supported during the period of this research initially by ASTRON
  grant 781-76-017 and subsequently by EC Marie Curie Fellowship
  ERBFMBICT 972436.  We thank Guy Pooley for making the radio
  observations for us, and acknowledge useful discussions with Andre
  van Teeseling, Peter Kahabka, Rob Hynes and Martin Still.
\end{acknowledgements}

\end{document}